\documentclass{desyproc}

\usepackage{amsmath,amssymb,bm}
\usepackage{graphicx}
\usepackage{amsfonts}

\begin{document}
%------------------------------------
\title{  XYZ mesons  and Glueballs: effects of the quark-gluon mixing }

%for single authors the superscripts are optional
\author{{\slshape Nikolai Kochelev }\\[1ex]
Institute of Modern
Physics, Chinese Academy of Science, Lanzhou 730000, China and \\
Bogoliubov Laboratory of Theoretical Physics, Joint Institute for Nuclear Research,\\
 Dubna, Moscow Region, 141980,  Russia }

% if the proceedings are available online (e.g. at Indico)
% please enter the contribution ID or file_name below for the DOI
%\contribID{32}
\contribID{smith\_joe}

% TO THE CONFERENCE EDITORS:
% please update the following information
% before sending the template to the authors
% \confID{800}  % if the conference is on Indico uncomment this line
\desyproc{DESY-PROC-2008-xx}
%\acronym{VIP08} % if you want the Acronym in the page footer uncomment this line
\doi  % if there is an online version we will register DOIs

\maketitle

\begin{abstract}
We present the new developments in the physics of XYZ mesons and glueballs and discuss the connection between properties of the $XYZ$ mesons and glueballs, and the structure of the QCD vacuum.
It is shown  that the  mixing between quark and gluon degree of freedoms induced by the nontrivial topological
structure of the QCD vacuum might affect strongly the properties of the exotic hadrons. A new interpretation of the  $XYZ$ mesons
as the mixing between radial excitations of light quark-antiquark systems and states with heavy quark content is suggested.
We introduce also a new mechanism of quarkonia production in hadron-hadron collisions coming from the mixing
of light and heavy quark sectors of  QCD. This mechanism is based on the nontrivial topological
structure of the QCD vacuum and might explain the polarization puzzle in inclusive $J/\Psi$ production in high energy reactions.
The possibility of big changes of the scalar and pseudoscalar glueball masses in  the hot  gluonic plasma is established and their consequences in the phase structure of QCD analyzed.

\end{abstract}

\section{Introduction}
Nowadays the study of the  exotic hadrons is  one of the hottest topics of hadron physics. Within the quark model the ordinary mesons  are considered as bound states of quark-antiquark pairs and baryons are  three-quark bound states. However, more complicated states which include additional quarks and/or valence
gluons are not forbidden by QCD.
There were  three big waves in the history of the exotic hadrons.
The first wave was created by the pioneering papers by Jaffe  \cite{Jaffe:1976ig,Jaffe:1976ih} on spectroscopy of
the four quark $q^2\bar q^2$  states within MIT bag model with light $u,d,s$ quarks. These states we call now light tetraquarks.
The second wave was created by the prediction of the $udud\bar s$ strange pentaquark state with very small width within the chiral soliton model
\cite{Diakonov:1997mm}. The third and last wave was  generated  by the Belle Collaboration discovery of the  $X(3872)$ meson with very small width and the mass near
$D D^* $ threshold \cite{Choi:2003ue}. Now we have many candidates for  exotic tetraquark states with heavy quark content
which are called the $XYZ$ mesons. There are also several candidates for the heavy pentaquarks (see, for example, the recent
reviews \cite{Olsen:2016hbz,Lebed:2016hpi}). A lot of the different microscopical models have been suggested to  explain their unusual properties.
We discuss a new approach to the $XYZ$ mesons based on the possibility of a large mixing between radial excitations in the
light quark and antiquark system
and states with heavy quark content. Such mixing might be  induced  by the nonperturbative quark-gluon interaction related to the nontrivial
topological structure of the QCD vacuum. The second part of this review is devoted to the modern status of
 glueballs which are the bound states of the gluons. We discuss the properties of  these exotic hadrons including the possibility of
 the existence of the three-gluon bound states. And finally, we show that glueballs can influence strongly the  properties of
 Quark-Gluon Plasma.

\section{The exotic XYZ mesons}
One of the main problems of the exotic hadron spectroscopy is the possibility of large mixings between exotic and nonexotic hadrons
with the same quantum numbers. This mixing does not allow,  in the most cases, to separate exotic states from the ordinary
 hadrons. The instantons, the strong fluctuations of the vacuum gluon fields,
 induce  t'Hooft's multiquark interaction
which can  lead to the strong mixing of the different quark flavors
 (see review \cite{Schafer:1996wv}).
Due to the specific structure of t'Hooft's interaction it can violate the OZI rule  in the channels with the  quantum numbers $0^{-+}$ and $0^{++}$ as shown in Fig.1a.
Since the instanton corresponds to a
subbarrier transition between vacua with the different topological
charge, the natural energy scale of this violation is
related to the height of the
potential barrier between these vacua. This height is given by the
energy of  the so-called sphaleron
$E_{sph}=3\pi/4\alpha_s(\rho_c)\rho_c \approx 2.8$  GeV with  $\rho_c\approx 0.3$~fm being the average instanton size in the QCD vacuum
\cite{Diakonov:2002fq}. For higher mass excitations  one can expect that the instanton-anti-instanton molecules  give an important contribution to the mixing between states with different quark flavors. Examples of such  mixing  between  light quark-antiquark states and  states with heavy quarks are shown in the Figs.1b, c and Fig.2.
The  nonperturbative interaction which governs that mixing  is  the quark-gluon effective interaction induced by instantons
\cite{Kochelev:1996pv,Kochelev:2009rf}
\begin{equation}
\mathcal{L}_I =- i\frac{g_s\mu_a}{4M_q} \, \bar q\sigma^{\mu\nu}t^a q \, G^{a}_{\mu\nu}
\label{Lag}
\end{equation}
where $G^{a}_{\mu\nu}$ is  the gluon field strength, $M_q$ is the effective mass of the light $u-, d-$   quarks in  the vacuum and
 $\mu_a$ is the anomalous quark chromomagnetic moment (AQCM).
\begin{figure}[h]
\centerline{\epsfig{file=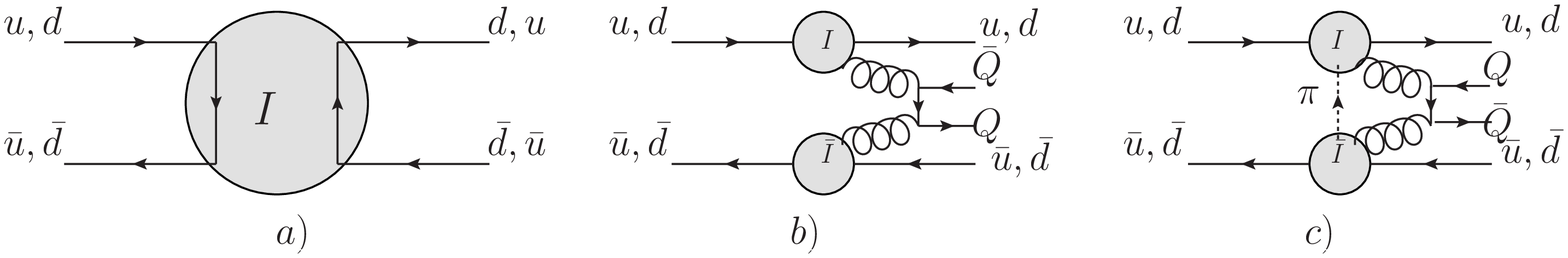,width=15cm,height=3cm, angle=0}}
\caption{ a) low energy mixing between light quarks induced by single instanton or anti-instanton , b)   high energy  mixing between light quarks
and heavy quarks $Q$ induced by the instanton-anti-instanton molecules and c) additional mixing induced by an additional pion exchange which is
shown by the dashed line.}
\end{figure}
\begin{figure}[h]
\centerline{\epsfig{file=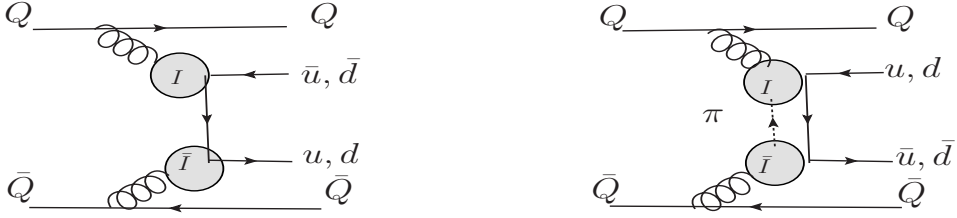,width=13cm,height=3cm, angle=0}}
\caption{ Additional mixing  between heavy and light quarks  in the  $I=0$ channel.}
\end{figure}
An extension of Eq.\ref{Lag}  which preserves the  chiral symmetry was suggested by Polyakov and
has the following form \cite{diakonov}
\begin{equation}
{\cal L}_I= -i\frac{g_s\mu_a}{4M_q}\bar q\sigma^{\mu\nu}t^a
e^{i\gamma_5\vec{\tau}\cdot\vec{\phi}_\pi/F_\pi}q G^{a}_{\mu\nu},
\label{Lag2}
\end{equation}
where  $F_\pi = 93~\mbox{MeV}$ is the pion decay constant \footnote{For simplicity we consider only the $N_F=2$ case.}.
 Within the instanton model, the value of AQCM is \cite{Kochelev:2009rf}
\begin{equation}
\mu_a=-\frac{3\pi (M_q\rho_c)^2}{4\alpha_s(\rho_c)}\approx -0.38
\label{AQCM1}
\end{equation}
for $M_q\approx 170$ MeV in mean field approximation.

\begin{figure}[h]
%\vskip -11cm
\centerline{\epsfig{file=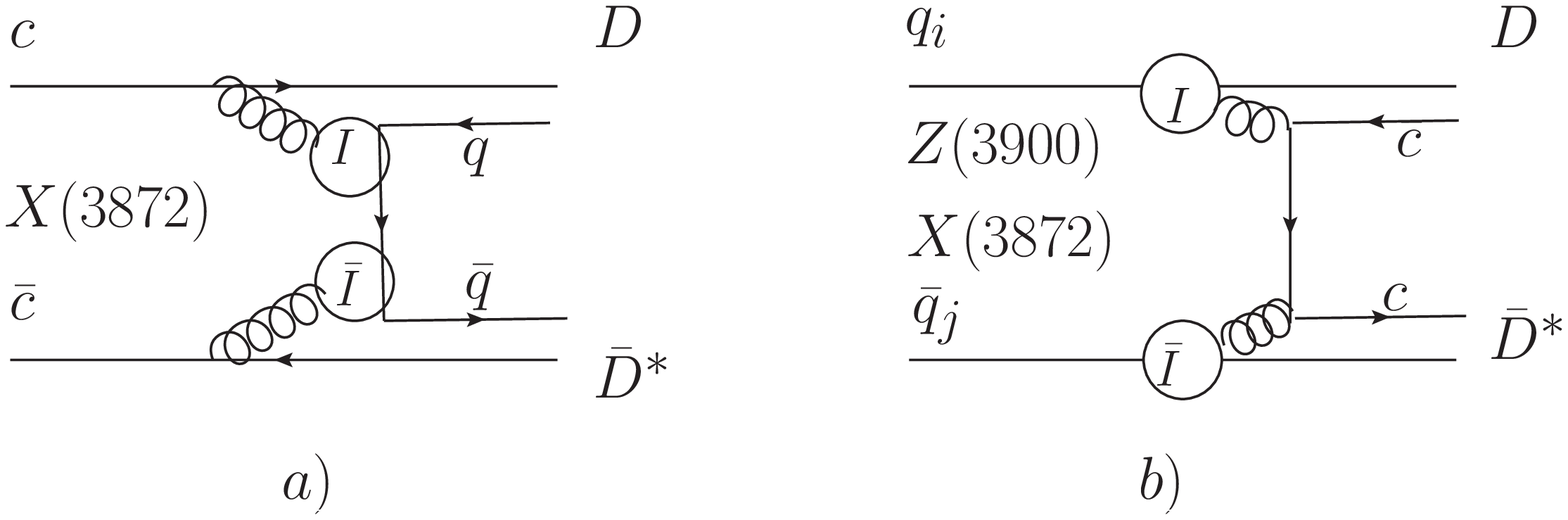,width=13cm,height=5cm, angle=0}}
%\vskip 12cm
\caption{ a)  Mixing $c\bar c$ core with $D \bar D^*$ state, b) mixing the light radial excited state with $D\bar D^*$   }
\end{figure}

In general, a  high radially excited state of a light quark-antiquark system can mix easily  with  states which have a
 heavy quark content, if they have the same quantum numbers  and  approximately the same mass.
It might be that the $X(3872)$ and the $Z_c(3900)$ are result of the mixing of molecular states and/or  tetraquark charmed hidden states with high radial excitations of a light quark system, by the mechanism shown in Fig.1b, c and Fig.3b.
 A recent paper in the spirit of this idea  was presented by  Coito \cite{Coito:2016ads}.
In the case of such mixing one should observe the decays of the XYZ mesons to final states without heavy quarks.
The absence of these decay modes in present experiments might be related to the
 existence of  many nodes in the wave function of the light quark system. One can estimate, for example,
  that light quark $q\bar q$ systems with radial numbers  $n_r \sim 7-8 $ have approximately the same mass as the $X(3872)$ and the $Z_c(3900)$.
 The many node structure of the wave function might lead to a very small overlap of the initial and final  wave functions in the decay.
 These strong effect of the many nodes in wave function in  decays is well known phenomenon.  In particular, Brodsky and Karliner in \cite{Brodsky:1997fj} used it to explain the small rate of $\Psi(2S)\rightarrow \rho\pi$  in  comparison with
 $J/\Psi\rightarrow \rho\pi$.
 Therefore, the mixing of high excited states in light quark-antiquark  systems
 with heavy quark states might be behind of the unusual properties of  XYZ mesons.
\section{ The $J/\Psi$ polarization puzzle}

There is a longstanding  problem to explain the absence of transverse polarization of $J/\Psi$  in the inclusive production
at high energy. The transverse  polarization was
predicted by perturbative QCD and it comes from  the  fragmentation  of on-shell gluon to charm-anticharm pair
(see review   \cite{Broadsky:2012rw}). Such polarization is in disagreement with the measurement
 at TEVATRON \cite{Abulencia:2007us} and at LHC \cite{Aaij:2013nlm, Chatrchyan:2013cla}.
 \begin{figure}[h]
\centering{\includegraphics[width=12cm,height=3cm,angle=0]{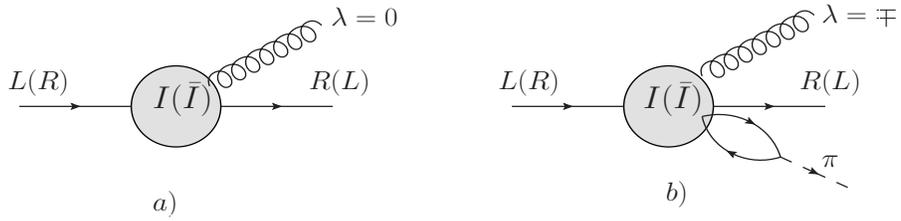}}
\caption{Diagram a) shows the  quark-gluon vertex generated by the  instanton,
 and diagram b) describes the quark-gluon-pion vertex induced by the  instanton.}
\label{fig:1}
\end{figure}
In a recent paper the polarization of gluons in the constituent quark  induced by instantons  was
calculated \cite{Kochelev:2015pqd}. The diagram without pion, Fig.4a, produces no polarization of the gluons \cite{diakonov}. However, the diagram
with an additional pion, Fig.4b,  leads to polarization of the gluons.  In this case the instanton and anti-instanton provide a
polarization of opposite sign. The contribution of the instantons to the $J/\Psi$ polarization in proton-proton collisions is presented in Fig.5.
It is evident, that such mechanism leads to a longitudinal polarization of  the $J/\Psi$ because the perturbative vertex with a hard gluon
does not change the helicity of the charm quark.
\begin{figure}[h]
\begin{minipage}[c]{6cm}
%\vskip -0.5cm\bar{\dot{}}
%\hspace*{1.0cm}
\centerline{\epsfig{file=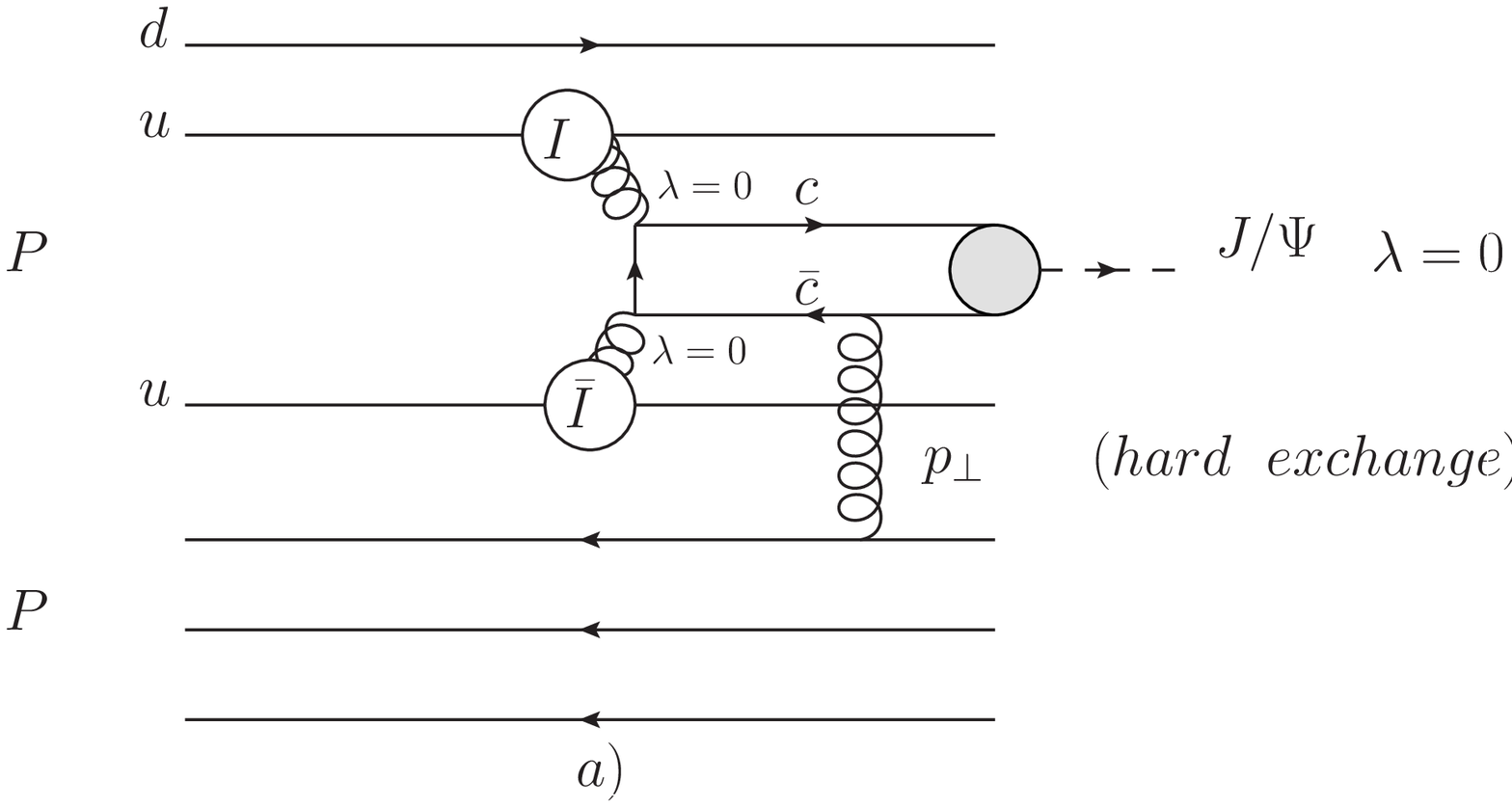,width=6.5cm,height=4.5cm, angle=0}}\
\end{minipage}
\begin{minipage}[c]{7cm}
\vskip -0.5cm
 \hspace*{1.5cm}
\centerline{\epsfig{file=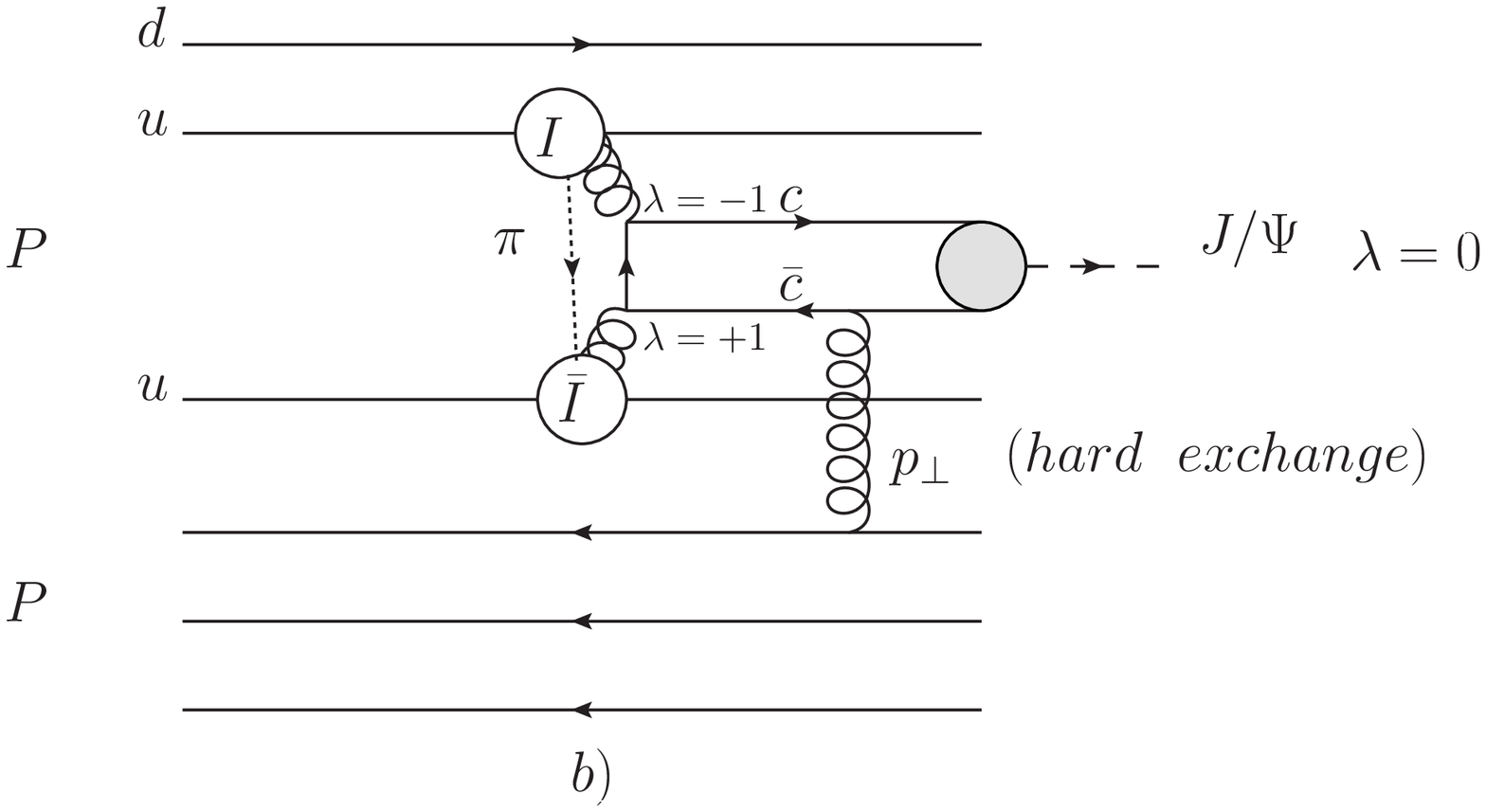,width=6.5cm,height=4.5cm, angle=0}}\
%\hspace*{1.0cm}
\end{minipage}
\caption{Example of diagrams which lead to the  longitudinal polarization of  $J/\Psi$  induced by instanton-anti-instanton pair a) without pion exchange and b) with pion exchange. }
\end{figure}
From the another side, one can consider this new mechanism as the contribution
to the
inclusive production of the quarkonia coming
from the mixing of light quarks with a state having  heavy quark content.
\section{Glueballs}
The glueballs, the bound states of the gluons, are now widely under discussion
(see the reviews \cite{Mathieu:2008me,Ochs:2013gi}). Different approaches including MIT bag model, several types of the constituent models,
holographic QCD,
QCD sum rule methods and the  lattice simulation were used to extract the information on the spectrum and quantum numbers of the glueballs.
One direct way based on the first principals of QCD to calculate the glueball spectrum are lattice calculations.
 At the present, the old  \cite{Morningstar:1999rf}  and more recent lattice calculations \cite{Chen:2005mg} in quenched QCD, i.e. the theory without quarks,
 are in the agreement and the  result for the low mass of the glueballs in pure $SU(3)_c$  is
 \begin{equation}
 M(0^{++})\approx \  1.7\  GeV,  M(2^{++})\approx \ 2.4\  GeV,
 M(0^{-+})\approx \ 2.6\ GeV \nonumber
 \end{equation}
 One of the interesting features of lattice calculations is that in the  unquenched QCD
 with  light quarks the result for the lowest mass of glueballs  practically does not change \cite{Gregory:2012hu}
  \begin{equation}
 M(0^{++})= 1.795(60)\  GeV, \ \  M(2^{++})= 2.620(50) \ GeV, \nonumber
 \end{equation}
 for 2+1 flavor and  $m_\pi=360$ MeV.
  A more  recent result for $N_f=2$  and $m_\pi\approx 580$ MeV  is \cite{Chen}
  \begin{equation}
 M(0^{++})= 1.624(141) GeV, M(0^{-+})= 2.738(153) GeV,  \  M(2^{++})= 2.516(95) GeV. \nonumber
 \end{equation}
%\begin{table}
There are many candidates for the glueball states.
For the  scalar glueballs we have the following mesons
\begin{equation}
f_0(600),\ \ f_0(980),\ \  \ f_0(1370),\ \ f_0(1500), \  \  f_0(1710),\  \   f_0(1790), \nonumber
 \end{equation}
  For
pseudoscalar glueball the candidates are
\begin{equation}
 \eta(1405), \  \   X(1835), \ \  X(2120), \ \  X(2370), \  \  X(2500)
\nonumber
\end{equation}
In the case of the tensor glueballs there are two candidates
\begin{equation}
 f_J(2220), \ \ f_2(2340).
\nonumber
\end{equation}
One of the main problems of the glueball spectroscopy is the possible large
 mixings of the glueballs with ordinary  and exotic  quark states
   which leads to the  difficulties  in disentangling the glueballs  in  the experiments (see recent discussion in \cite{Vento:2015yja}).
The example of such mixing  is presented in Fig.6.
\begin{figure}[h]
%\vskip -11cm
\centerline{\epsfig{file=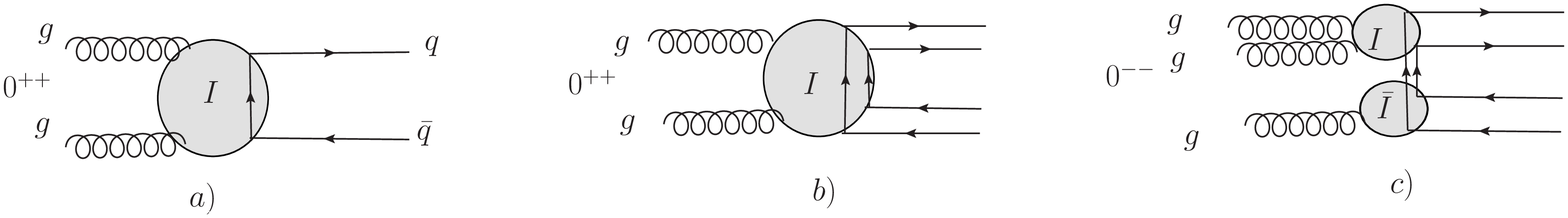,width=14cm,height=4cm, angle=0}}
%\vskip 12cm
\caption{ a)  Mixing $0^{++}$ glueball with a quark-antiquark state, b) mixing  of this glueball with a tetraquark, and c)
mixing of the exotic three gluon $0^{--}$ state with an exotic tetraquark.}
\end{figure}
The contribution  of the mixing presented in Fig.6a was considered in \cite{Harnett:2008cw} within the QCD sum rule method, and it  was shown
to be very important.
One of the suitable ways to avoid this problem and obtain a clear prediction for the glueballs is to study the glueballs with exotic quantum numbers.
$J^{PC}=0^{--},0^{+-},1^{-+},2^{+-}, etc \ldots$ which cannot mix with ordinary quark-antiquark states.
 However, the calculation of the masses and decay modes of these exotic glueballs is very difficult  because one should
 consider states with a large number of gluons.
For example, Lattice calculations give a  very large mass $ M_G=5.17$ GeV  with a large uncertainty  for  the $0^{--} $
exotic three-gluon glueball.
The first attempt to calculate the masses of the glueballs with exotic quantum number  was done in ref. \cite{Qiao:2014vva,Qiao:2015iea}
within the QCD sum rule approach and a rather small mass for $0^{--}$ glueball was found.
However in a recent paper \cite{Pimikov:2016pag}, it was shown that the interpolation current for the gluonic state used in  \cite{Qiao:2014vva,Qiao:2015iea}
has an anomaly and cannot be  applied  for  the glueball calculations. In ref. \cite{Pimikov:2016pag}  a new interpolating current for
the $0^{--}$ glueball has been suggested.  The derived QCD sum rules are consistent and stable.
The resulting value for the mass  of the $0^{--}$ glueball state is
$M_G=6.3^{+0.8}_{-1.1}$ GeV and  an upper limit for the total width $\Gamma_G\leq 235$~MeV has been  obtained.
It has been also argued that the mixing of this glueball state with $0^{--}$  tetraquark is expected to be very small.
Therefore, the exotic $0^{--}$ glueball can be considered  a pure gluon state.
At the present, the Belle Collaboration is searching  for this state  \cite{Jia:2016cgl}.
\section{Glueballs in Quark-Gluon Plasma}
In  experiments at  RHIC and LHC a new type of nuclear
matter the so-called strongly interacted quark-gluon
plasma has been discovered (see review \cite{Shuryak:2014zxa}).
 It behaves as a liquid  and does not have the expected gas-like behavior.
There are many experiments planned at the running and future facilities to investigate the properties
of such new matter.
It is very important  from the point of
view of theory of the strong interactions  to find the fundamental  reason which leads to such
behavior of quark-gluon matter.  One possible way is to investigatethe simpler case of pure Gluodynamics, the theory without quarks.
Recently  very  precise lattice results for the Equation of State (EoS) of this
theory at finite $T$ below and above the deconfinement temperature $T_c$
were presented \cite{Borsanyi:2012ve}. They challenge our understanding of QCD dynamics at finite temperatures.
One of the puzzles is a very spectacular  behavior of the trace anomaly,
$I/T^4=(\epsilon-3p)/T^4$,
as a function of $T$. In particular, just above $T_c$ it rapidly grows untill
$T_G\approx 1.1 T_c$ and then it decays  as $I/T^4\sim 1/T^2$ untill $T\approx 5T_c$.
A mechanism which can explain  such anomalous behavior has been suggested in ref. \cite{Kochelev:2016hrx}.
 The mechanism is based on the possibility  of a large change of the glueball masses above  the $T_c$.
The  starting point is the relation between the lowest scalar glueball mass, $m_G$, and the gluon condensate,
$G^2=<0|~\frac{\alpha_s}{\pi} G_{\mu\nu}^a G_{\mu\nu}^a|0>$ at  $T=0$,
which  appears naturally in the dilaton approach \cite{Lanik}
\begin{equation}
m_G^2f_G^2=\frac{11N_c}{6}<0|\frac{\alpha_s}{\pi}G_{\mu\nu}^a G_{\mu\nu}^a|0>,
\label{dilaton}
\end{equation}
 where  $f_G$ is the glueball coupling constant to gluons. Lattice calculations show that the gluon
condensate decreases roughly by a factor two at  $T=T_c$  due to the
strong suppression of its electric component, while slightly above  $T_c$ the condensate vanishes
very rapidly due to the cancellation between its magnetic and electric components \cite{Lee:2008xp}. The temperature behavior of the condensate at
 $T_G\geq T \geq T_c$ can be described by the equation \cite{Miller:2006hr}

\begin{equation}
G^2(T)=G^2\bigg[1-\bigg(\frac{T}{T_G}\bigg)^n\bigg]
\nonumber
\end{equation}
 and
 the mass of scalar and pseudoscalar glueballs for  $T_G\geq T \geq  T_c$ are given by
\begin{equation}
m(T)=m_0\sqrt{1-\bigg(\frac{T}{T_G}\bigg)^4}.
\nonumber
\end{equation}
The possible dynamical reason for the decrease of scalar and pseudoscalar glueball masses above $T_c$
is the strong attraction between gluons induced by instanton-antiinstanton molecules shown in Fig.7.
Such a molecular structure of the instanton vacuum is expected above $T_c$
  For the quark-antiquark channel the effect of the instanton-antiinstanton induced interaction
was discussed in \cite{brown} and it was shown to be strong.
For the gluon-gluon interaction the effect should be even stronger because it is enhanced by factor $S_0^2\approx 10^2$, where
$S_0\approx 10$ is the instanton action.
\begin{figure}[htb]
\centerline{\epsfig{file=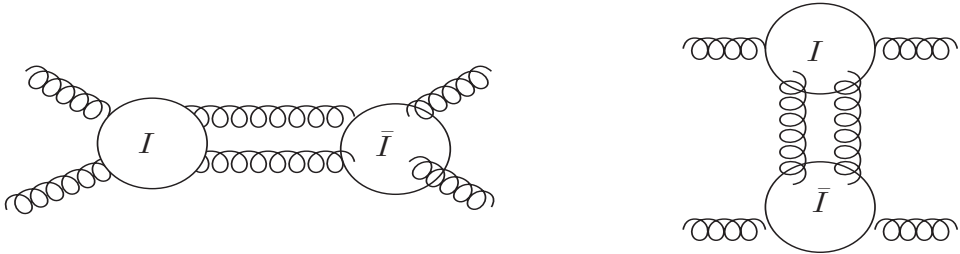,width=13cm,height=3.5cm, angle=0}}
\caption{Gluon-gluon interaction induced by instanton-antiinstanton molecules.}
\end{figure}
\begin{figure}[h]
\centerline{\epsfig{file=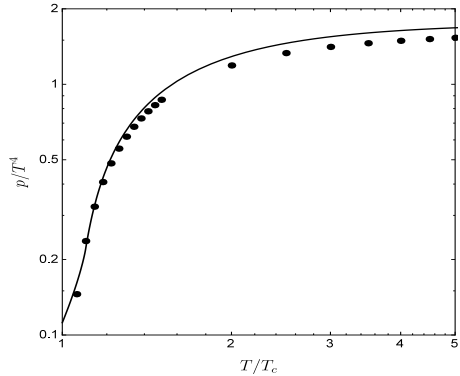,width=6cm,height=5cm, angle=0}}\
\caption{ The pressure $p/T^4$ as the function of $T/T_c$  in comparison with the  lattice data \cite{Borsanyi:2012ve}.}
\end{figure}
The rapid decrease  of the scalar and preudoscalar glueball  masses above $T_c$ leads to a large contribution of these
glueballs to the pressure of the gluon plasma and the sum of the glueball and free gluon contributions describe the lattice
  data very well as shown in Fig.8. Given these results we can describe three different phases of gluon matter  \cite{Kochelev:2016hrx}.
   The first phase is the confinement regime at  $T<T_c$ and is described by a gas of  massive glueballs.
The second phase appears   just above  $T_c$ at   $T_G\geq T\geq T_c $,
 where $T_G\approx 1.1 T_c$. In this phase
 the main contribution to the EoS is coming from  the very light scalar and pseudoscalar glueballs.
  Above the $T_G\approx 1.1T_c$ the third phase appears,  a mixed phase of massless gluons and point-like  scalar-pseudoscalar massless  glueballs.
 The existence of such phases can be checked in high multiplicity events in relativistic heavy ion collisions where
  an abundant production of glueballs is expected \cite{vento,glueballs}. The most simple way is to look for the change
  of the scalar and pseudoscalar glueball masses in the  two pion and three pion channels, correspondingly, as a function of
  multiplicity. Finally it should be  also mentioned that the mixing between quark and gluon states can  change strongly above $T_c$ due to the
  transition between single instanton induced mixing to the mixing produced by instanton-antiinstanton molecules.

\section{Conclusion}

The  exotic hadrons carry a very  important
information on the structure of the strong interactions and
the properties of the exotic hadrons are very sensitive to the nonperturbative  structure
of  QCD.
 We have shown that the   XYZ mesons can arise as
 the result of a nonperturbative  mixing  between different configurations with and without
heavy quark content. We have furthermore suggested a new mechanism for heavy quarkonia production at high energy which can be related
to such mixing. We have also discussed the modern status of the glueballs and their mixing with quarkonia. Finally, we have predicted a drastic change of the scalar and
pseudoscalar glueball masses above the deconfinement temperature in the gluonic plasma  which has very exciting consequences at the level of the possible phase transitions at very high temperatures.

\section*{Acknowledgments}
I would like to thank Alexander Dorokhov, Sergo Gerasimov, Mikhail Ivanov, Ernst-Michael Ilgenfritz,
   Makoto Oka, Pengming Zhang, Bing-Song Zou
 and, especially,  Vicente Vento  for
useful discussions. I am grateful to Ahmed Ali and Michail Ivanov for the invitation to give a talk about exotic hadrons at 
Helmholtz - DIAS International Summer School
"Quantum Field Theory at the Limits:
from Strong Fields to Heavy Quarks", 18-30 July 2016, Dubna, Russia.
This work has been supported by the
  Chinese Academy of Sciences President's International Fellowship Initiative
  (Grant No. 2013T2J0011), the Japan Society for the Promotion
of Science (Grant No.S16019) and was initiated through the series of APCTP-BLTP JINR Joint Workshops.

\end{document}